\documentclass[%
 aip,
 amsmath,amssymb,
 reprint,%
]{revtex4-1}

\usepackage{graphicx}
\usepackage{dcolumn}
\usepackage{bm}

\usepackage[utf8]{inputenc}
\usepackage[T1]{fontenc}
\usepackage{mathptmx}
\usepackage{etoolbox}

\makeatletter
\def\@email#1#2{%
 \endgroup
 \patchcmd{\titleblock@produce}
  {\frontmatter@RRAPformat}
  {\frontmatter@RRAPformat{\produce@RRAP{*#1\href{mailto:#2}{#2}}}\frontmatter@RRAPformat}
  {}{}
}%
\makeatother
\begin{document}

\preprint{AIP/123-QED}

\title[Hybridization, degeneracy and polarization independence of  modes in square array of graphene disk resonators with symmetry $\bf C_{4v}$]{Hybridization, degeneracy and polarization independence of modes in square array of graphene disk resonators with symmetry $\bf C_{4v}$}
\author{Victor Dmitriev}
 \email{victor@ufpa.br}
\author{Amanda Evangelista da Silva}
 \email{amanda.evangelista.silva@itec.ufpa.br}
\affiliation{
Faculty of Electrical and Biomedical Engineering,
Institute of Technology, Federal University of Pará, Belém 66075-110, Brazil
}%

\date{\today}

\begin{abstract}
 We consider an infinite array of graphene disks in a square unit cell, where every disk of the quadrumer is excited in the dipole regime. Hybridization of plasmonic eigenmodes belonging to different IRREPs of the~point group of symmetry $C_{4v}$ is discussed. It is shown that the hybridization of the pair of 1D IRREPs modes $A_1$ and $B_2$ as well as the pair $A_2$ and $B_1$ can exist in this symmetry. For the cell period $p_0$ and the quadrumer period $P$, the relation $P/p_{0}=2$ defines polarization independence of these pairs of modes. The discussed modes are dark ones and require for their excitation breaking the point symmetry of the quadrumer and, for breaking their degeneracy, i.e. separation in the two pairs, a modification of the discrete translation symmetry, i.e. change the~parameter $P/p_{0}$. The separated modes can be excited by normally incident plane wave with orthogonal polarizations.
\end{abstract}

\maketitle


In resonator theory of electromagnetic, acoustic, mechanical  and other systems, when different eigenfunctions have the same resonant frequency, they are called degenerate. Usually, when discussing possible modes in a system, one relates degeneracy to the dimensionality of the irreducible representations (IRREPs) of the point group describing  symmetry of the system \cite{overvig2020selection}. For example, the two-dimensional (2D) IRREP $E$ of the point symmetry $C_{4v}$, describing a square unit cell in an electromagnetic array, defines  polarization degeneracy of two orthogonal dipole modes. The cube resonator is described by the octahedral point group $O_{h}$, which has 3D IRREPs. Therefore, eigenmodes of such resonator can be triply degenerate. These degeneracies are symmetry defined.

Another type of degeneracy, which is not related to symmetry, namely, accidental degeneracy, can occur in a resonator system for some combination of physical and geometrical parameters. An example of such degeneracy is presented in \cite{Transition}. Both types of degeneracy, symmetry-based and accidental one can be lifted by introducing an internal perturbation or by applying an external field.

Because of the  losses in the graphene disks which provide existence of surface plasmon-polariton  (SPP) modes, our system is  non-Hermitian, therefore, the eigenmodes are not orthogonal and the eigenfrequencies are complex. In this case, strictly speaking, one can not use the theory of IRREPs. As a physical approximation, we shall consider the modes as ``quasi-orthogonal", neglecting losses. This allows one a systematic classification of the modes basing on their symmetry.

In this Letter, we discuss  hybridization of modes and a specific degeneracy which is related to symmetry, however, not with the dimensionality of the corresponding IRREPs. 
We will show that, in an array with $C_{4v}$ symmetry, eigenmodes belonging to~one-dimensional IRREPs $A_1$, $B_2$, $A_2$ and $B_1$ can be degenerate by symmetry in a special manner. In some way, this degeneracy is hidden. It concerns the symmetry-protected (dark) modes and can be discovered by applying a perturbation. When discussing such degeneracy in periodic systems, one should take into account both the point symmetry of the unit cell and the discrete translational symmetry of the array. 

\begin{figure}[ht]
\centering
\includegraphics[width=0.9\linewidth]{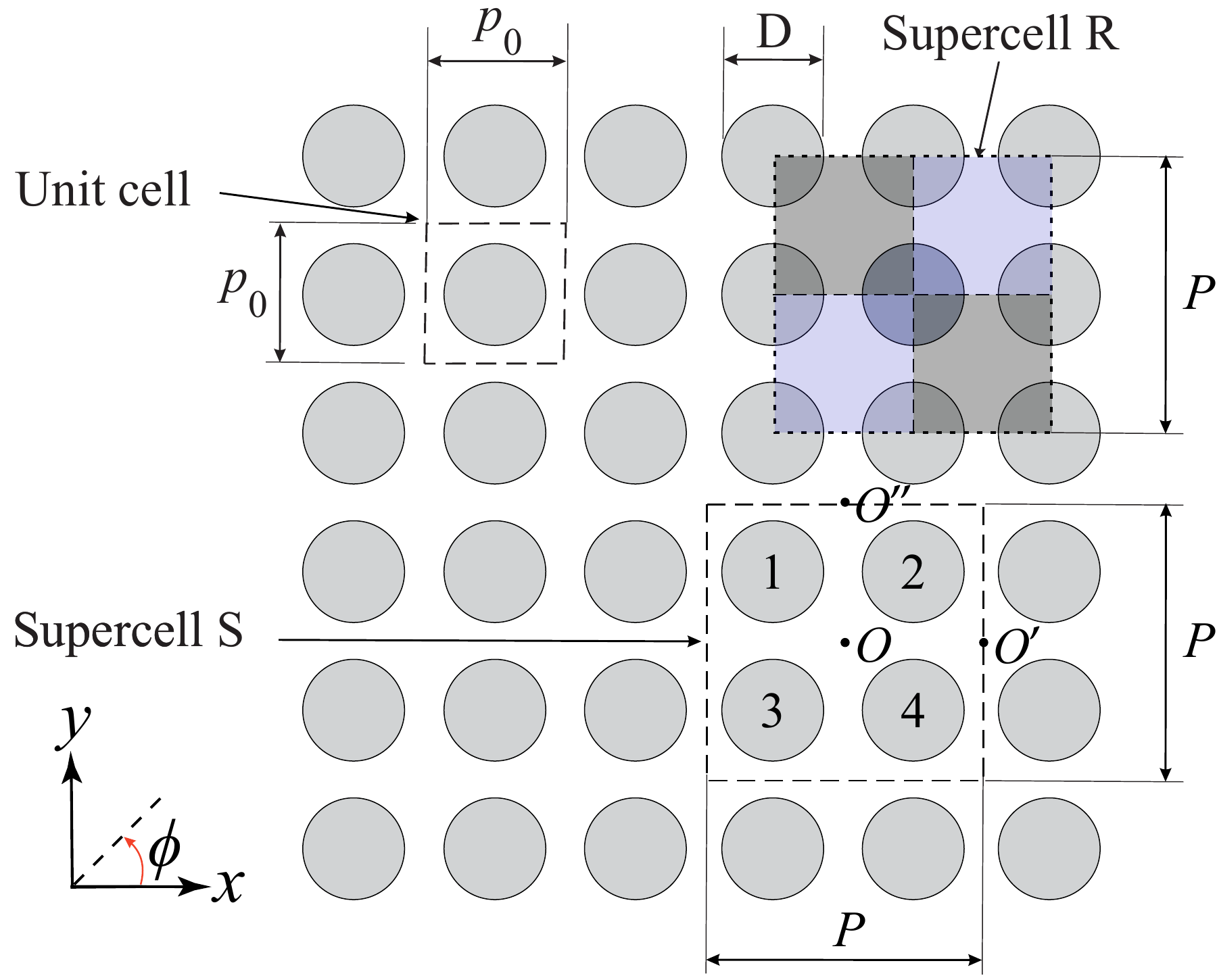}
\caption{\label{fig:Arrays}Schematic of a free-standing array composed by graphene disks with diameter $D$. A square unit cell with one disk and the period $p_0$ and a square supercell $S$ with four disks and with the period $P=2p_0$ are marked by dashed lines. Another possible square supercell $R$ with a disk at its center is shown by dotted lines.}

\end{figure}

%
 As an illustrative example, we have chosen an infinite electromagnetic array of graphene disks presented schematically in Fig.~\ref{fig:Arrays}. An~individual graphene disk can support SPP modes with strong confinement of~ electromagnetic field \cite{wang2012edge}. Arrays with such disks have also been  investigated (see, for example, \cite{deop2002optical,dmitriev2025thz}).

 This subwavelength array is a lattice with the period $p_0$, where the disks have the diameter $D$. For simplicity, we consider a free-standing array. The presence of a dielectric substrate will not change the principal results of the theory. For the following discussion, we choose, firstly,  a supercell $S$ with four disks (i. e. quadrumer), marked in Fig.~\ref{fig:Arrays} by a dashed square. The period of the quadrumer is $P=2p_0$. 
 
 Now, considering the quadrumer as a translation unit, one can choose its center at the point $O$ or at the point $O'$, dislocated by half of the quadrumer period $P$ in the $x$-direction (see Fig.~\ref{fig:Arrays}). For an infinite periodic system, it does not matter which of these two points to choose in the analysis, because they are equivalent. The reference point also can be displaced by $P/2$ in the $y$-direction to the equivalent point $O''$.

In THz region, the optical response of graphene is determined by intraband transitions, which can be modeled using the Kubo formalism \cite{gonccalves2016introduction}. The surface conductivity $\sigma(\omega)$ depends on~the~angular frequency $\omega$ and the chemical potential~$\mu_{c}$ as~follows:
\begin{eqnarray}
\sigma(\omega) = \frac{\sigma_{0}}{\pi}\frac{4}{\hbar(\gamma - i\omega)} \left[ \mu_{c} + 2k_{B}T \ln \left( 1 + e^{-\frac{\mu_{c}}{k_{B}T}} \right) \right]
\label{eq:Kubo_full},
\end{eqnarray}
where $\sigma_{0} = e^{2}/4\hbar$ represents the universal conductivity, $e$~is the electron charge, $\hbar$ is the~reduced Planck’s constant, $k_{B}$ is~the~Boltzmann constant, $\gamma = 1/\tau$ denotes the scattering rate with $\tau$ being the carrier relaxation time. In our examples below, we assume $T = 300$~K, $\tau = 1$~ps and, for the nonperturbed array, $\mu_{c}=1$~eV.

As a first step in the analysis, we apply to the eigenmodes of~the~supercell $S$ in Fig.~\ref{fig:Arrays}. All of the eigenmodes are based on the dipole mode of an individual disk. The disks are relatively close to one another, so that coupling between the adjacent disks by near electromagnetic fields provide existence in the quadrumer eigenmodes with the symmetry $C_{4v}$. Therefore, the resonant frequencies of these eigenmodes are localized not very far from the dipole resonance of the individual disk.

For the numerical example, we choose the following parameters of the array: $P=2p_0= 15$ \textmu m, $D=6$ \textmu m. The condition $P/p_0=2$ corresponds to a uniform distribution of the disks in the $x$0$y$ plane and will be modified in some calculus. The electromagnetic simulations were fulfilled by software COMSOL Multiphysics. The calculated parameters and characteristics of the array eigenmodes are presented in Table~\ref{tab:Dipoles} and Table~\ref{tab:Higher}.

\begin{table}[ht]
\caption{\label{tab:Dipoles} Dipole eigenmodes (IRREPs  $\mathbf{E}$): SALC images, $E_z$ fields, currents, resonant frequencies, and $Q$-factors.}
\begin{ruledtabular}
\begin{tabular}{cccc}
\textbf{IRREP} & 
\textbf{\begin{tabular}[c]{@{}c@{}}SALC\\ image\end{tabular}} & 
\textbf{\begin{tabular}[c]{@{}c@{}}Currents\\ and fields\end{tabular}} & 
\textbf{\begin{tabular}[c]{@{}c@{}}Frequency (THz),\\ $\bf \textit{Q}$-factor\end{tabular}} \\
\hline
$E_{\Gamma}$ & 
$\vcenter{\hbox{\begin{tabular}{c}
\includegraphics[width=2.1cm]{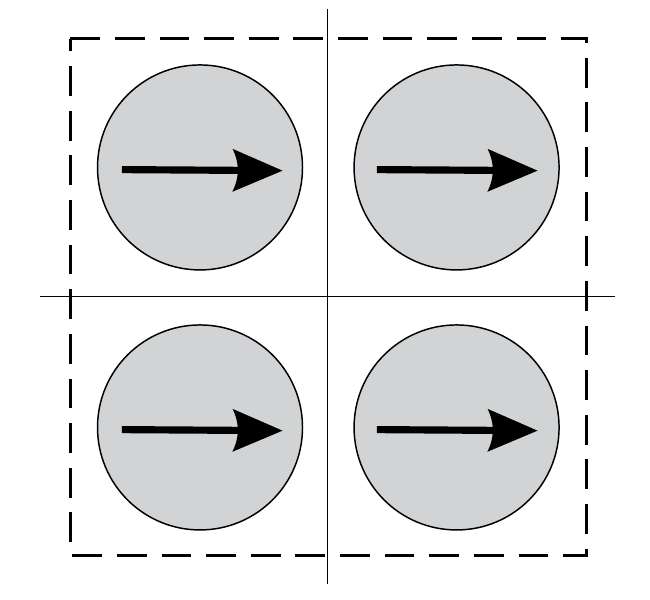}\\ \vspace{0.1cm}
\includegraphics[width=2.1cm]{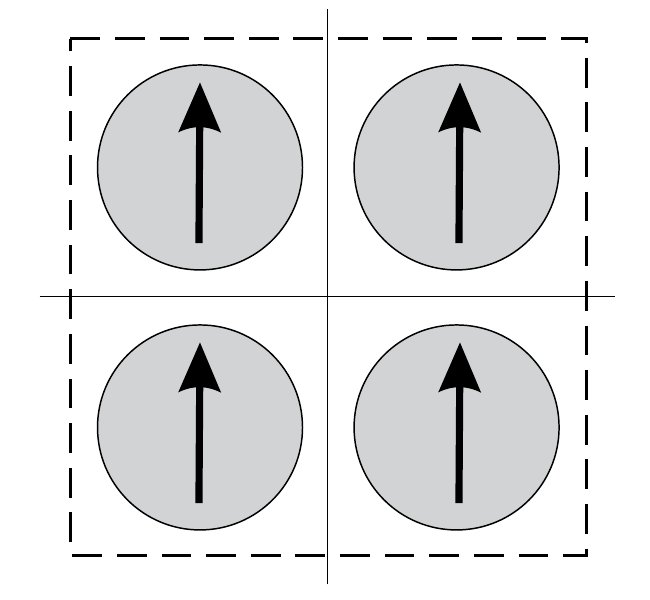}
\end{tabular}}}$ & 
$\vcenter{\hbox{\begin{tabular}{c}
\includegraphics[width=2.1cm]{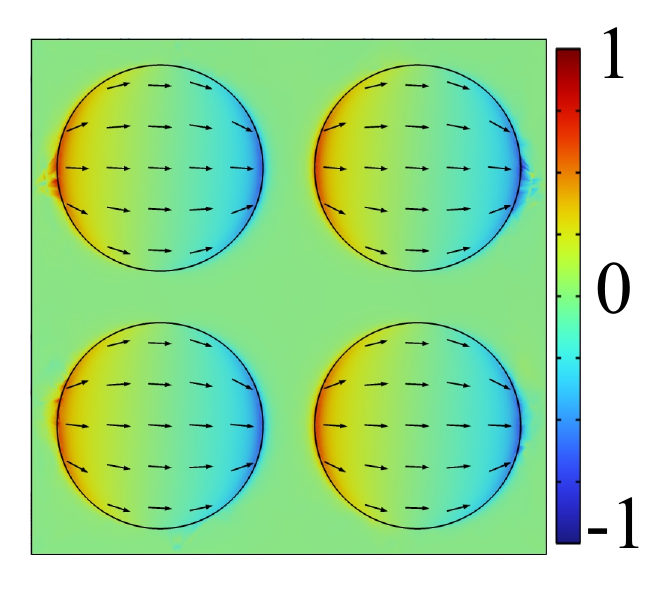}\\ \vspace{0.1cm}
\includegraphics[width=2.1cm]{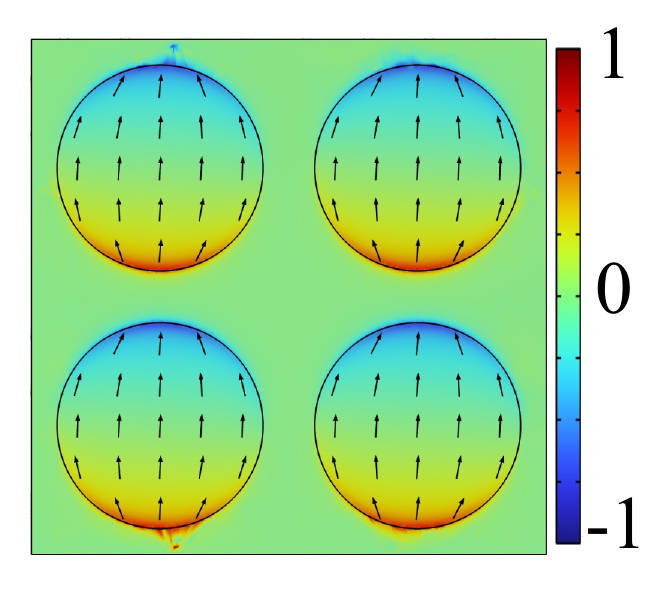}
\end{tabular}}}$ & 
$\mbox{\begin{tabular}{c}
Dipole $D_{x}$\\
$(7.001+0.8806i)$\\
$Q = 3.97$ \\[1cm]
Dipole $D_{y}$ \\
$(7.005+0.8802i)$ \\
$Q = 3.98$ 
\end{tabular}}$\\
\hline
$E_{X}$ & 
$\vcenter{\hbox{\begin{tabular}{c}
\includegraphics[width=2.1cm]{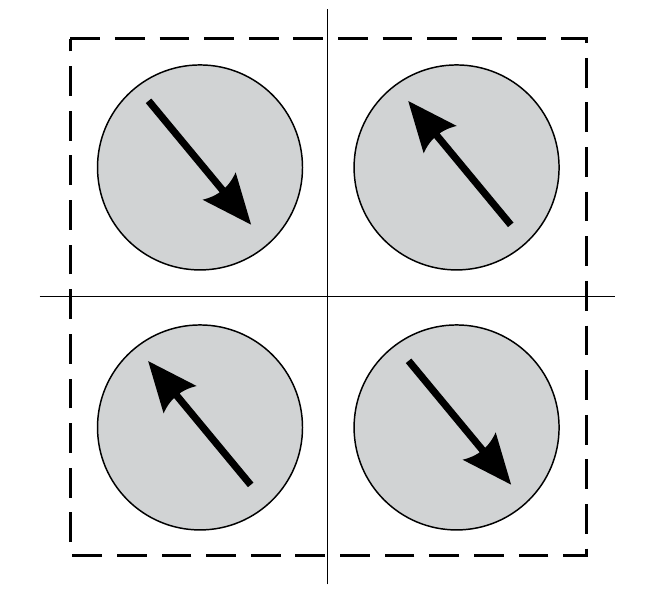}\\ \vspace{0.1cm}
\includegraphics[width=2.1cm]{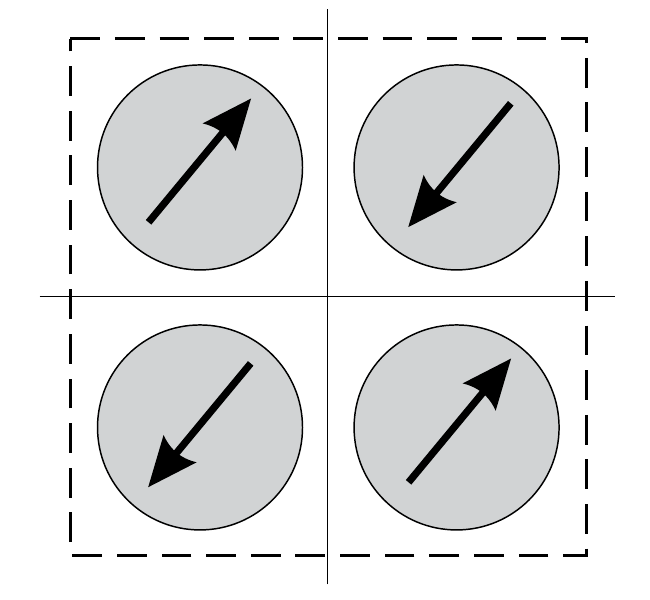}
\end{tabular}}}$ & 
$\vcenter{\hbox{\begin{tabular}{c}
\includegraphics[width=2.1cm]{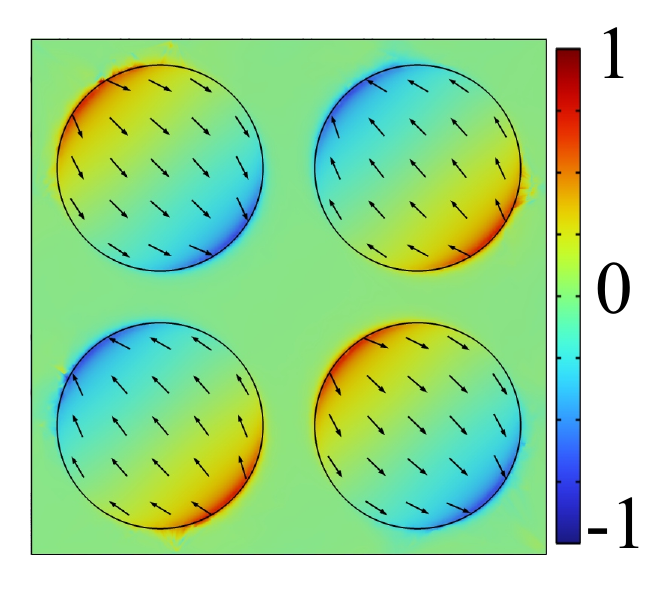}\\ \vspace{0.1cm}
\includegraphics[width=2.1cm]{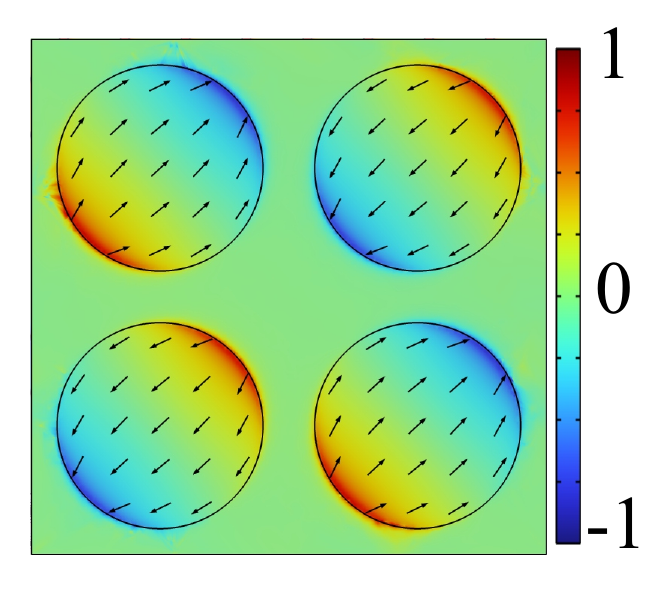}
\end{tabular}}}$ & 
$\mbox{\begin{tabular}{c}
Dipole $D_{-xy}$\\
$(7.664+0.0840i)$\\
$Q = 45.59$ \\[1cm]
Dipole $D_{xy}$\\
$(7.670+0.0841i)$ \\
$Q = 45.59$
\end{tabular}}$ \\
\end{tabular}
\end{ruledtabular}
\end{table}

\begin{figure}[htbp]
\centering
\includegraphics[width=0.8\linewidth]{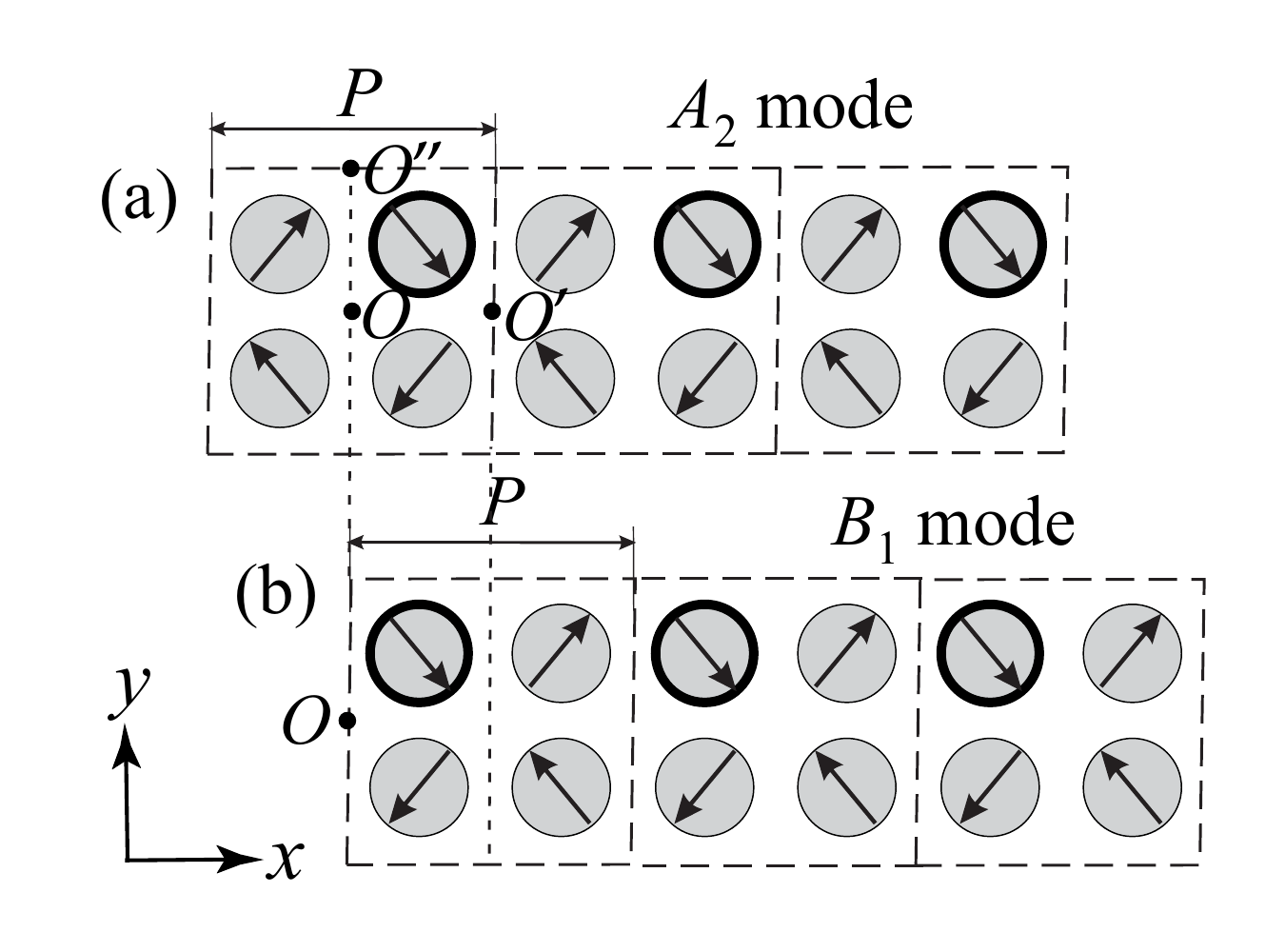}\\
\includegraphics[width=0.8\linewidth]{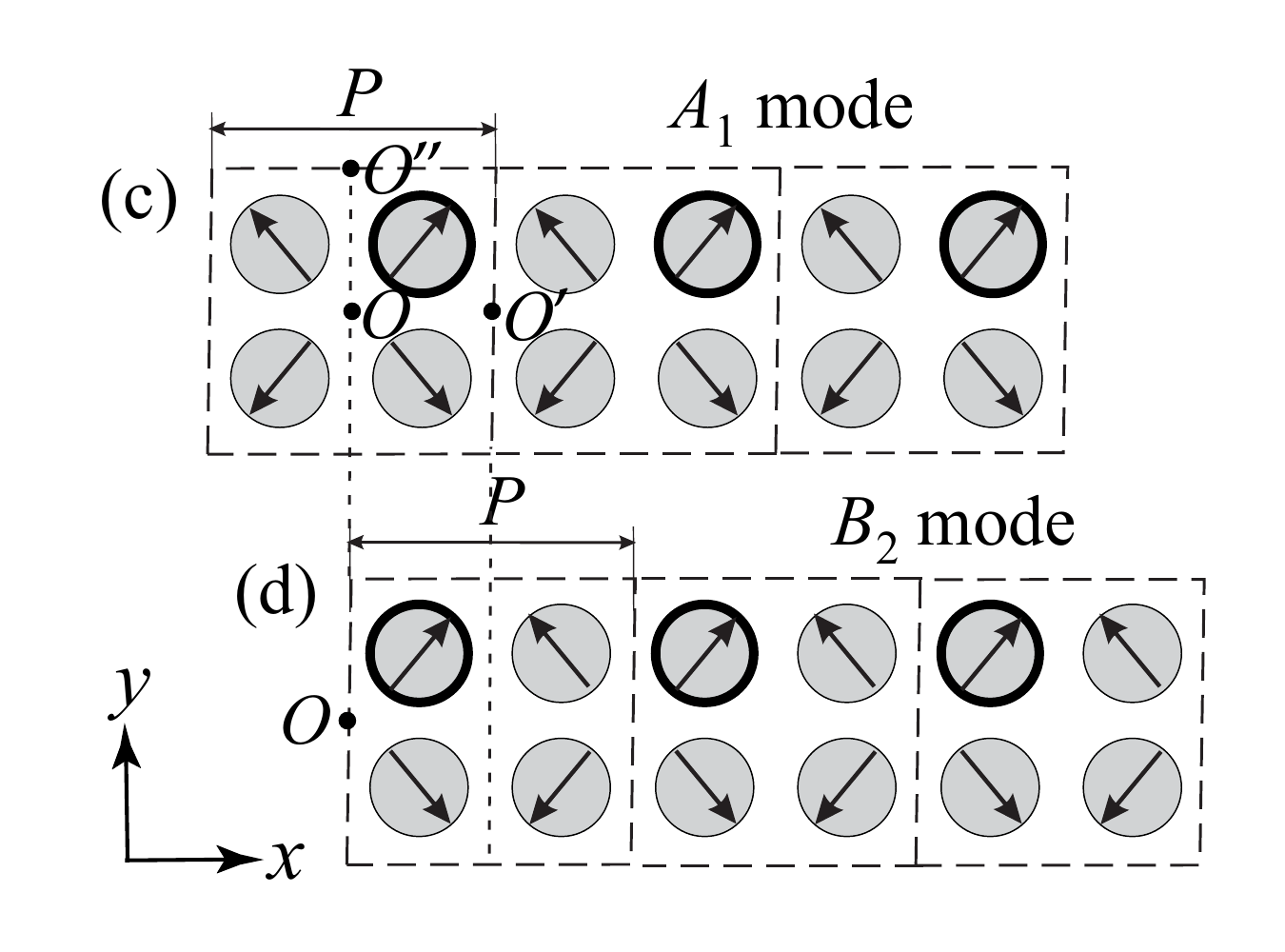}
\caption{\label{A2B1_A1B2}SALC image of three periods of the mode (a) $A_2$, (b)~$B_1$~obtained by dislocation of $A_2$ along the $x$-axis by half of the period $P$. 
(c) $A_1$, (d) $B_2$ obtained by dislocation of $A_1$ along the $x$-axis by half of the period $P$. Supercell $S$, perturbed disks are denoted by bold circles.}
\end{figure}
\begin{table}[ht]
\caption{\label{tab:Higher}$A_1$, $A_2$, $B_1$, and $B_2$ IRREP eigenmodes: SALC images, $E_z$ fields, currents, resonant frequencies, and $Q$-factors.}
\begin{ruledtabular}
\begin{tabular}{cccc}
\textbf{IRREP} & 
\textbf{\begin{tabular}[c]{@{}c@{}}SALC\\ image\end{tabular}} & 
\textbf{\begin{tabular}[c]{@{}c@{}}Currents\\ and fields\end{tabular}} & 
\textbf{\begin{tabular}[c]{@{}c@{}}Frequency (THz),\\ $\bf \textit{Q}$-factor\end{tabular}} \\
\hline

$A_{1}$ & 
$\vcenter{\hbox{\includegraphics[width=2.1cm]{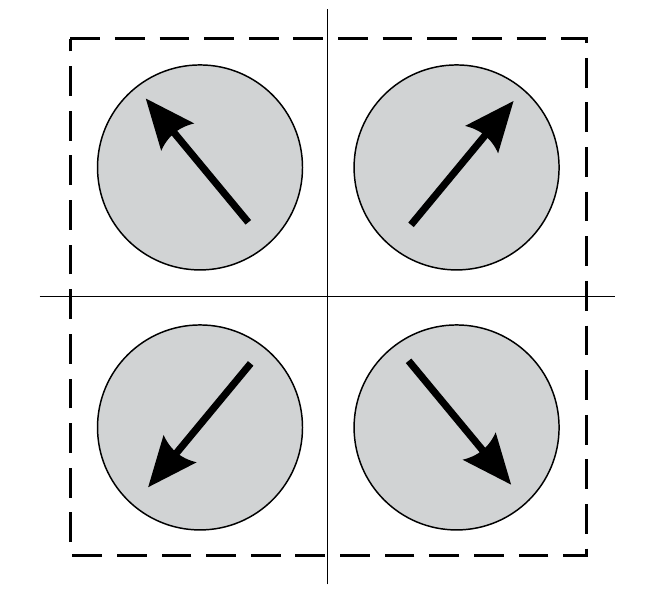}}}$ & 
$\vcenter{\hbox{\includegraphics[width=2.1cm]{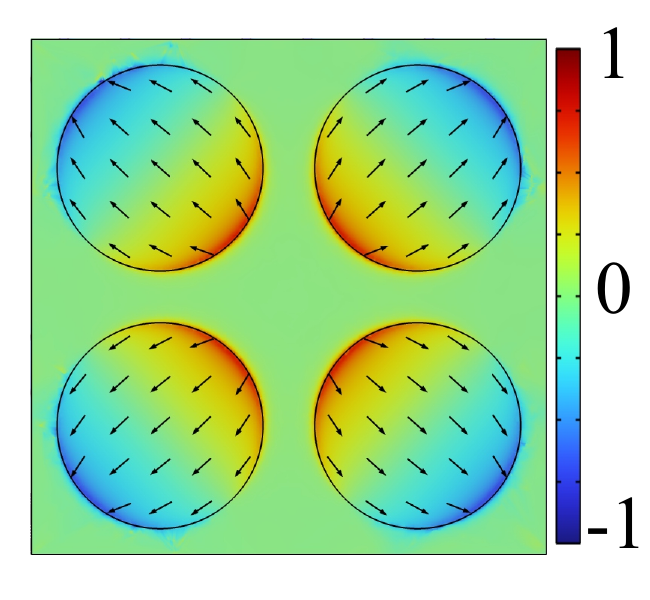}}}$ & 
$\vcenter{\hbox{\begin{tabular}{c} $(8.223+0.0932i)$ \\ $Q = 44.14$ \end{tabular}}}$ \\
\hline

$A_{2}$ & 
$\vcenter{\hbox{\includegraphics[width=2.1cm]{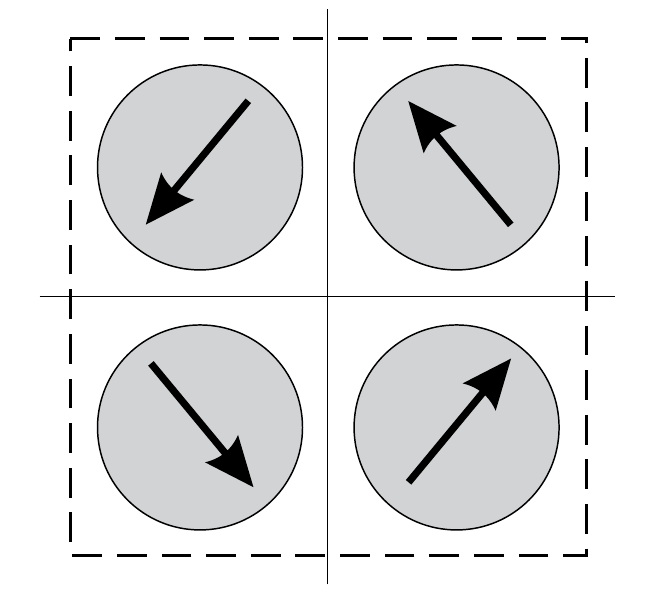}}}$ & 
$\vcenter{\hbox{\includegraphics[width=2.1cm]{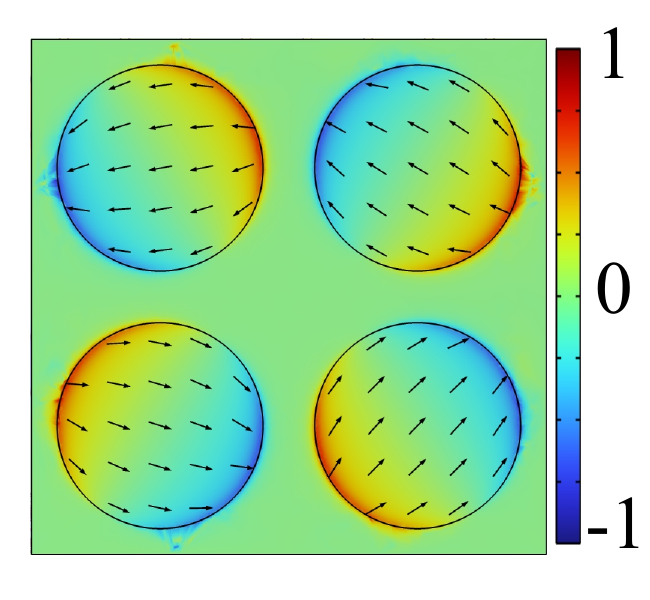}}}$ & 
$\vcenter{\hbox{\begin{tabular}{c} $(6.653+0.0724i)$ \\ $Q = 45.95$ \end{tabular}}}$ \\ 
\hline

$B_{1}$ & 
$\vcenter{\hbox{\includegraphics[width=2.1cm]{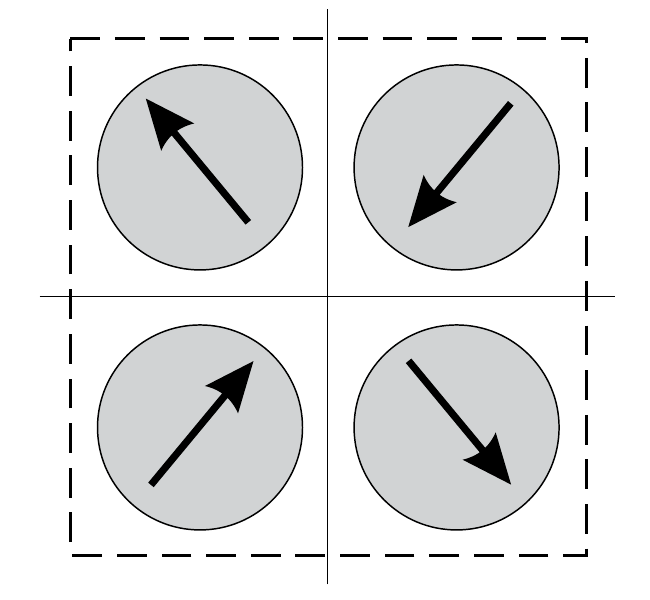}}}$ & 
$\vcenter{\hbox{\includegraphics[width=2.1cm]{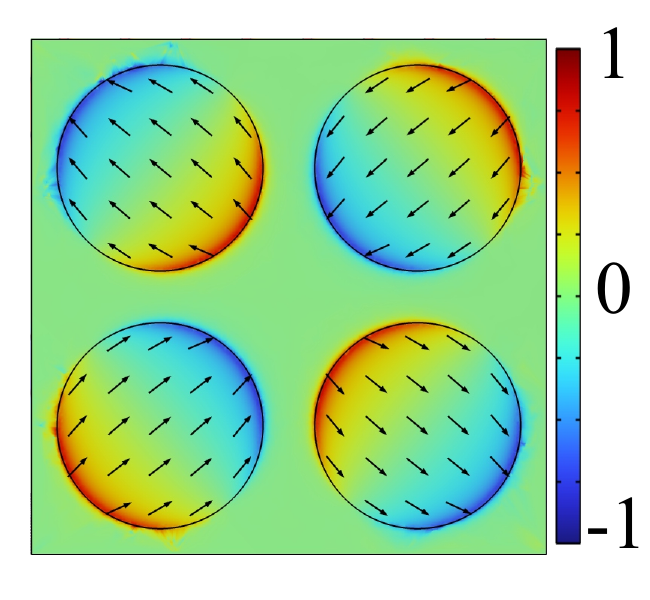}}}$ & 
$\vcenter{\hbox{\begin{tabular}{c} $(6.661+0.0724i)$ \\ $Q = 45.98$ \end{tabular}}}$ \\
\hline

$B_{2}$ & 
$\vcenter{\hbox{\includegraphics[width=2.1cm]{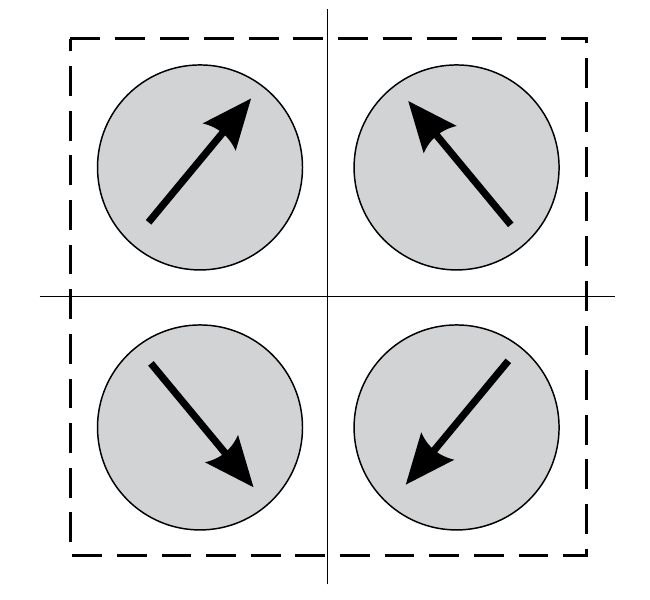}}}$ & 
$\vcenter{\hbox{\includegraphics[width=2.1cm]{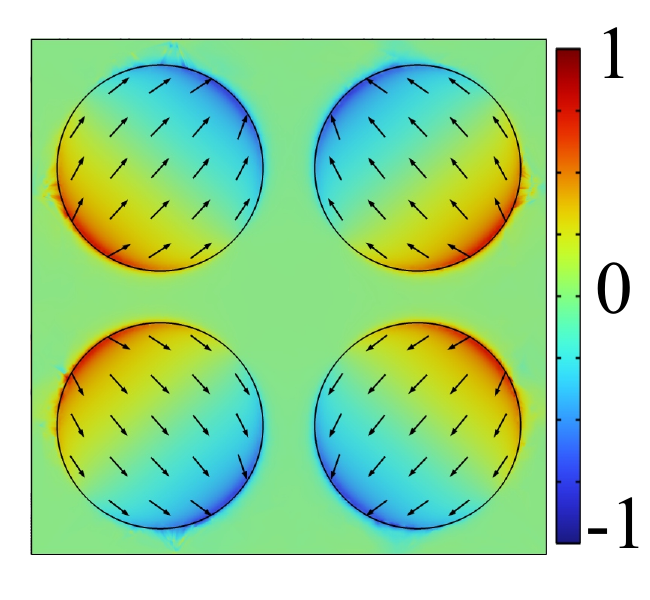}}}$ & 
$\vcenter{\hbox{\begin{tabular}{c} $(8.223+0.0931i)$ \\ $Q = 44.16$ \end{tabular}}}$ \\
\end{tabular}
\end{ruledtabular}
\end{table}

In proximity of the resonant frequency of the ordinary degenerate dipole SPP modes belonging to $\Gamma$ point of the first Brillouin zone (FBZ) of the group $C_{4v}$ (Table~\ref{tab:Dipoles}, first line, 2D IRREP $E_\Gamma$), another degenerate pair of modes exists. It belongs to $X$-point of the FBZ of the group $C_{4v}$~\cite{overvig2020selection}. The SALC (symmetry adapted linear combination \cite{dmitriev2025thz}) images of the two diagonal partners of this mode are shown in Table~\ref{tab:Dipoles} in the second line (2D IRREP $E_X$). We can call this a ``compensated'' dipole mode, because any two adjacent graphene dipoles in the quadrumer are oriented anti-parallel and, therefore, the net dipole moment of the supercell $S$ is zero. It is a symmetry-protected (dark) mode. A reduction in the symmetry of the quadrumer $S$ from $C_{4v}$ to $C^d_{s}$ (``$d$'' means diagonal) by a perturbation of disk 2, for example, transforms this mode in quasi-dark one with a possibility of excitation. The two partners of the mode $D_{xy}$ and $D_{-xy}$ can be excited by incident plane wave with diagonal polarization of electric field~$E_{xy}$ ($\phi=45^o$) or~$E_{-xy}$ ($\phi=135^o$), respectively. 


Next, we will show that in the discussed array can exist a specific degeneracy which is related to symmetry but it is not associated with the dimensionality of the IRREPs. This concerns a combination of two 1D IRREP modes, namely, the pair $A_1$ and $B_2$ modes and also the pair $A_2$ and $B_1$ modes, shown in Table~\ref{tab:Higher}. It is known that a linear combination of any two eigenmodes is also an eigenmode of the array. The degeneracy of the modes $A_2$ and $B_1$ can be explained qualitatively using SALC images in Fig.~\ref{A2B1_A1B2}(a)-(b) and for modes
$A_1$ and $B_2$ in Fig.~\ref{A2B1_A1B2}(c-d). In both tables, arrows denote currents and color maps define $E_z$ electric fields. 

The analyzed array is infinite, therefore, the choice of the center of the quadrumer is arbitrary. If we choose the center at point $O$ (Fig.~\ref{A2B1_A1B2}(c)), the dipoles in the quadrumer correspond to the mode $A_1$ (see~Table~\ref{tab:Higher}). But choosing the reference point at point $O'$, translated in $x$-direction by $P/2$ (Fig.~\ref{A2B1_A1B2}(d)), one deals with the mode $B_2$. The modes $A_1$ and $B_2$ have the same even parity with respect to the planes of symmetry $\sigma_x$ and $\sigma_y$, and the same odd parity with respect to the planes $\sigma_{xy}$ and $\sigma_{-xy}$. Matching of angular $\phi$ distribution of the quadrupole mode $B_2$ and the radial mode $A_1$ is provided by the field independence of $\phi$ in the mode $A_1$. Besides, half of the mode $A_1$ and half of the mode $B_2$ coincide due to the partial superposition of the quadrumers associated with these modes (see Fig.~\ref{A2B1_A1B2}(c-d). The other half of the mode $B_2$ can be obtained by dislocation the corresponding half of $A_1$ by the period $P$. The same effect exists for translation of the center $O$ by a half of the period $P$ in $y$-direction of the array. The specific structure of resulting fields is defined by the hybridization of the two modes. The eigenfrequencies of these modes coincide (see Table~\ref{tab:Higher}), therefore they are degenerate.

 Now we apply to excitation of the discussed modes by normally incident plane wave, breaking the $C_{4v}$ symmetry by a perturbation. This can be achieved by different methods, for example, by changing the diameter of one disk. In our example, the reduction of symmetry is fulfilled by changing, via doping, the chemical potential of the upper right disk in the quadrumer in Fig.~\ref{A2B1_A1B2}(c) (disk 2 in Fig.~\ref{fig:Arrays}) from $\mu_{c2}=1.0$~eV to $\mu_{c2}=0.5$~eV. This reduces the symmetry of the supercell $S$ from $C_{4v}$ to $C^d_{s}$ and defines the reference points and the reference planes of symmetry in the array. Perturbation reduces the resonant frequency of the perturbed disk and results in a lower amplitude of the disk dipole current and, consequently, in an imbalance of the current distribution in the four disks of the quadrumer. As a consequence, a net electric dipole moment, oriented in the diagonal direction appears in the quadrumer and this provides coupling of the array with external excitation~$E_{xy}$ or $E_{-xy}$.

\begin{figure}
\centering
\includegraphics[width=0.8\linewidth]{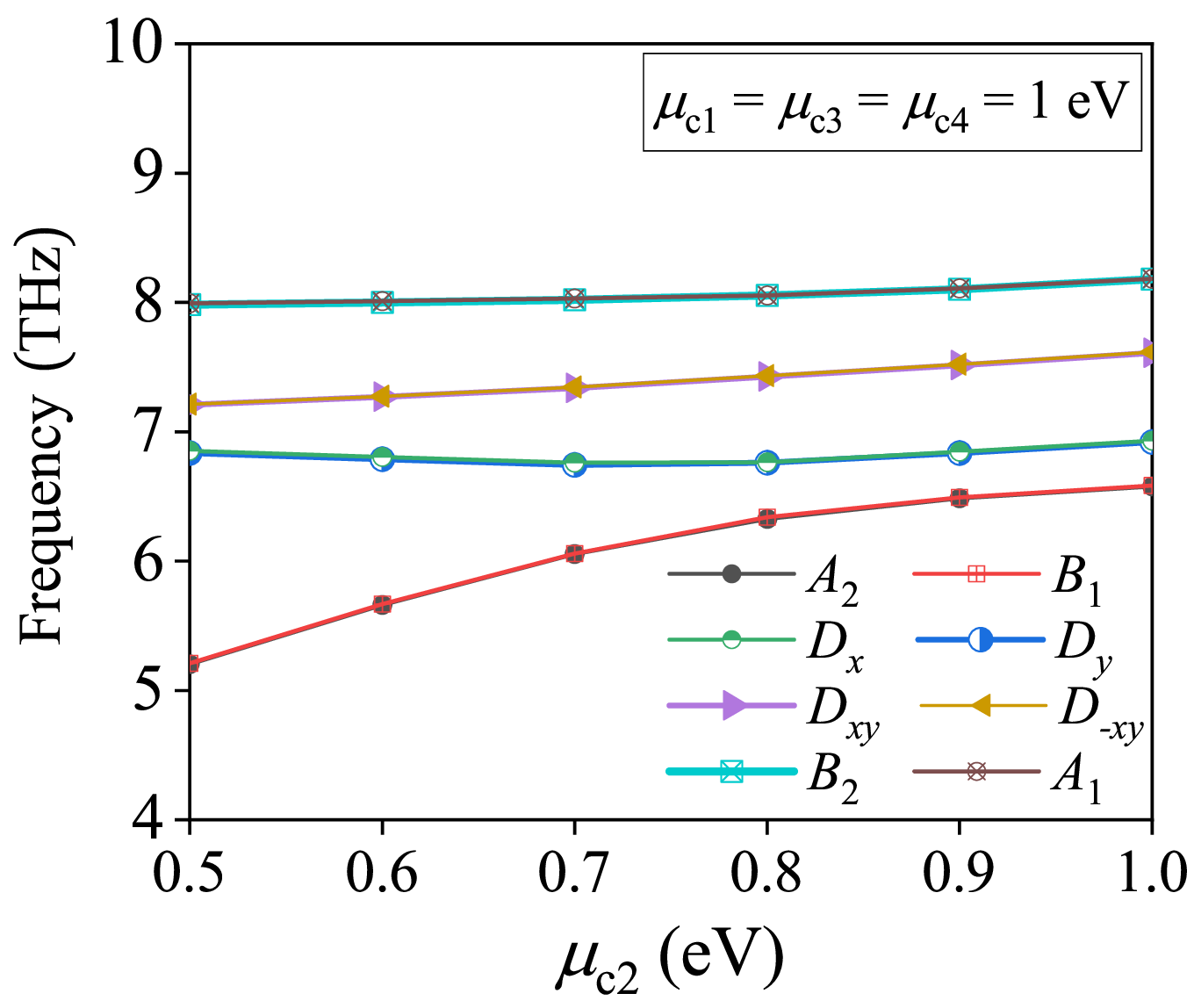}
\caption{Resonant frequencies of the eigenmodes versus chemical potential $\mu_{c2}$ of the disk 2 in Fig.~\ref{fig:Arrays},  $P/p_{0}=2$.}
\label{fig:mmucBvariation}
\end{figure}

Analyzing the current distribution of the mode ($A_1$, $B_2$) in Fig.~\ref{A2B1_A1B2}(c-d), one can see that the effect of the perturbation is  equal for both modes because the perturbed disk belongs to the hybrid pair ($A_1,B_2$). It can be seen also considering the square supercell $R$ in Fig.~\ref{fig:Arrays} with the perturbed disk in the center of the square. Symmetry of this supercell $R$ is $C_{4v}$. When a perturbation is applied, the resonant frequencies of both $A_1$ and $B_2$ modes will change, and the perturbation affects equally on these modes, independently on the level of perturbation. This is confirmed by numerical modeling in Fig.~\ref{fig:mmucBvariation} where the pairs ($A_1,B_2$) and ($A_2,B_1$) do not split with changing the potential $\mu_{c2}$.

Symmetry of the hybridized pairs ($A_1$, $B_2$) and ($A_2$, $B_1$)  can be discussed in terms of the wallpaper group $p4m$ \cite{overvig2020selection} (the chessboard pattern) and also in terms of the color groups \cite{dmitriev2025thz} (in our case, it is the group $C_{4v}(C_{2v})$). The rotational symmetry of ($A_1$, $B_2$) mode in the supercell $R$ is $C_2$, and, formally, one cannot expect polarization independence in the array. However, in the longwave approximation, the far-field properties of the infinite chessboard pattern with supercell $R$ do not change if we rotate the structure by $90^{\circ}$. Therefore, one can observe polarization independence of the transmittance spectra  in  Fig.~\ref{fig:mmucB=0,5eV} for both the pair ($A_1$, $B_2$) and the pair ($A_2$, $B_1$), in agreement with the group-theoretical predictions. 

\begin{figure}
\centering
\includegraphics[width=0.8\linewidth]{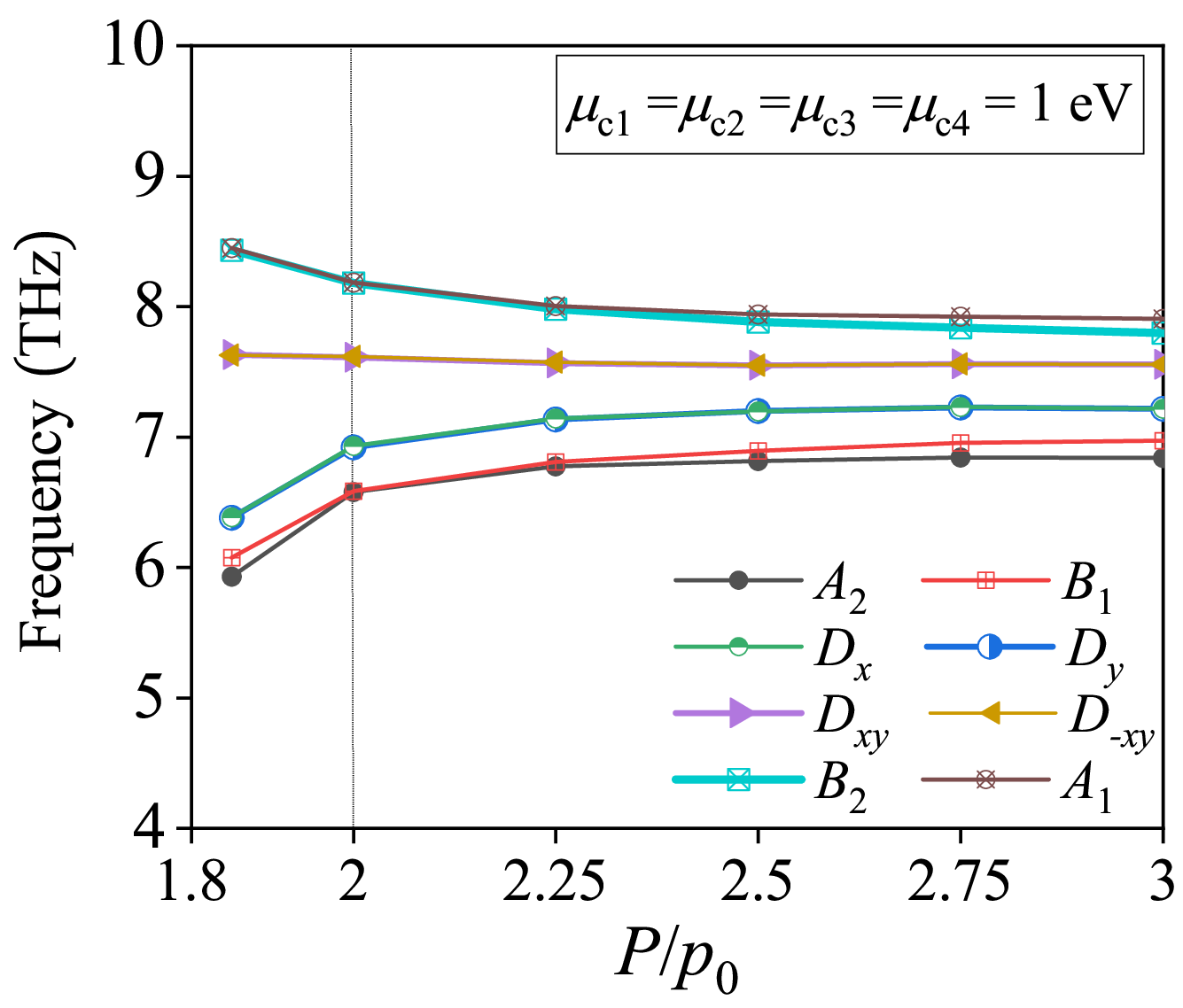}
\caption{Splitting of the resonant frequencies of ($A_1$, $B_2$) modes and of ($A_2$, $B_1$) modes due to changing translational symmetry of the array by the parameter $P/p_0$, $p_{0}=7.5$ \textmu m.}
\label{fig:Splitting}
\end{figure}

If we separate or approximate the quadrumers, choosing the parameter $P/p_0 \neq 2$, the discussed pairs of modes will be separated in frequency. These results for eigenwaves are shown in Fig.~\ref{fig:Splitting}. Thus, by modifying the translational symmetry, one can split the pair ($A_1$, $B_2$) and also the pair ($A_2$, $B_1$). Each of the two modes in the split pair can be excited by one of the two   orthogonally polarized waves.

In the calculated transmittance spectrum in Fig.~\ref{fig:mmucB=0,5eV}, small discrepancies in the resonant frequencies for different polarizations in the cases of curves (III) and (IV) can be explained by a violation of the longwave approximation at higher frequencies. The first resonance (dip I) in this figure corresponds to ($A_2$, $B_1$) pair, the second one (dip II) to the $E_\Gamma$ mode, the third resonance (dip III) to the $E_X$ dipole, and the fourth one (dip IV) to the pair ($A_1$, $B_2$). The resonance of the nonperturbed array with ordinary dipole $E_\Gamma$ mode is shown by dotted line. The chemical potential of the four disks for the last case was taken in calculus as arithmetic mean of the potentials of four disks in the perturbed quadrumer, i.e. $\mu_{c}=0.875$ eV. 

One can see in Fig.~\ref{fig:mmucB=0,5eV}, that the $Q$-factors of the quasi-BIC modes are one order higher than that of the ordinary dipole mode $E_\Gamma$. High value of $Q$-factor of the modes and a possibility of their dynamic control by chemical potential using a gate voltage, the specific electromagnetic field distribution of the modes and a possibility of controlling the resonances by polarization of external excitation can be useful for sensing, polarization conversion, switching and filtering purposes.

\begin{figure}[htbp]
\centering
 \includegraphics[width=0.8\linewidth]{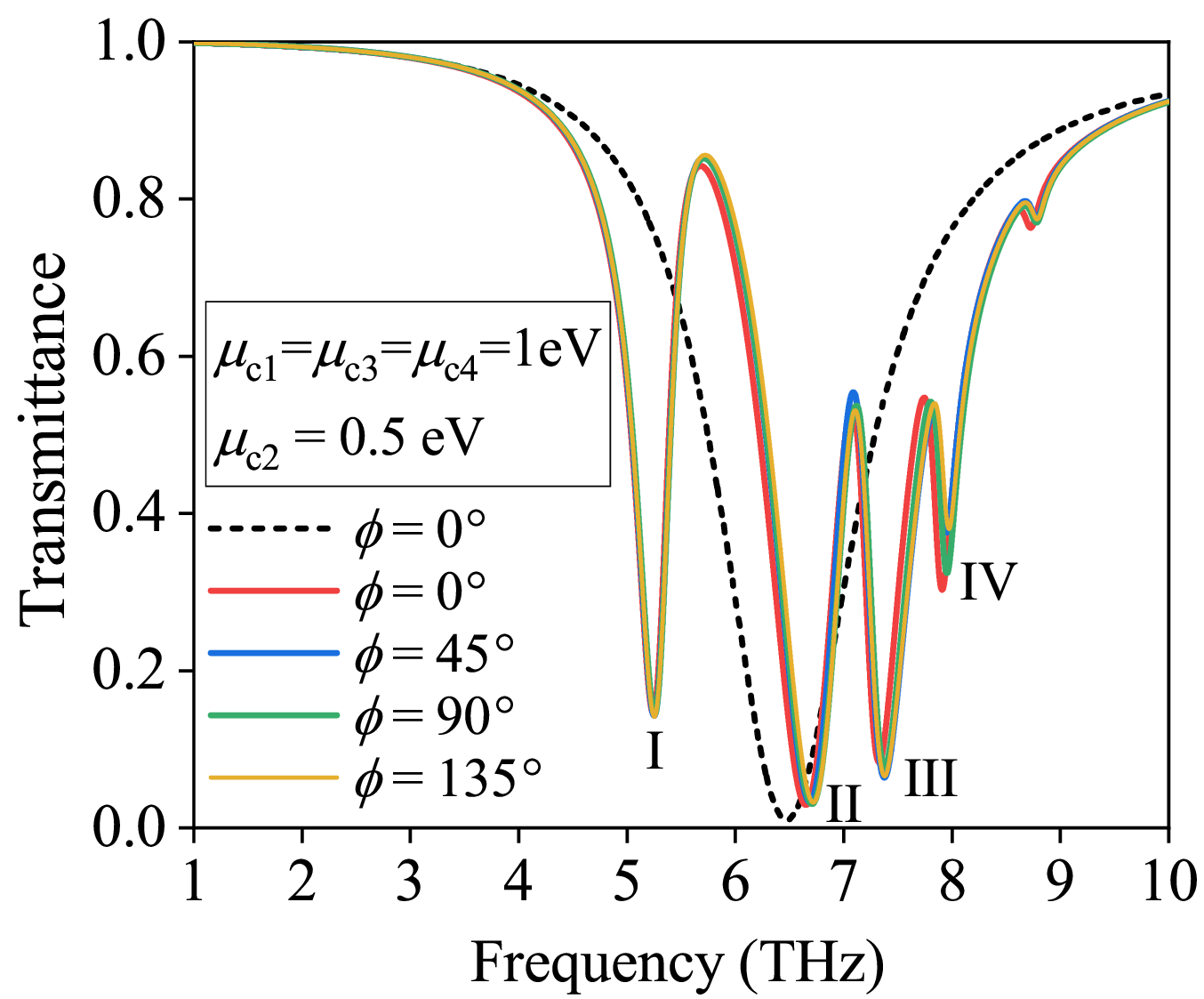}
\caption{Frequency characteristics of array for different polarization  angles $\phi$, $p_0=7.5$~\textmu m, $P/p_{0}=2$.}
\label{fig:mmucB=0,5eV}
\end{figure}

The discussed properties of the arrays are based on symmetry arguments, therefore, the results are quite general. These results are invariant under a change of the disks diameter and the array period (preserving the condition $P/p_0 = 2$). The results also do not depend on the material of the resonators, which in 3D structures can be dielectric, magnetodielectric, or metallic ones. In the analysis of such resonators by SALC method, instead of electric currents, in-plane electric or magnetic field vectors can be used. 

Summarizing, we have shown that in a square array of graphene disk quadrumers with $C_{4v}$ symmetry, hybridization and degeneracy of dark modes can exist. In order to excite the discussed modes separately, one should perform two types of perturbations. Firstly, reduction of the point symmetry of the quadrumer can provide access to the dark degenerate modes. As a result, in the transmission spectra of the array, besides the  dipole modes, two resonances corresponding to the pairs of the eigenmodes ($A_1$, $B_2$) and ($A_2$, $B_1$) will appear. Secondly, in order to separate two degenerate eigenmodes in every pair, one should break (better to say, modify) the translational symmetry of the initial array and this causes splitting of the resonant frequencies of the two modes in each pair. Thus, we can consider the discussed properties of the modes as a consequence of symmetry. The group-theoretical predictions are confirmed by analytical SALC method and by numerical calculations.

\begin{acknowledgments}
We wish to acknowledge the National Council for Scientific and Technological Development (CNPq) and the Amazonia Foundation for Support of Studies and Research (FAPESPA) for the financial support.
\end{acknowledgments}

\section*{Data Availability Statement}

Data available on request from the
authors.


\bibliography{aipsamp}
\end{document}